# Estimation of Physical Activity Level and Ambient Condition Thresholds for Respiratory Health using Smartphone Sensors

Chinazunwa Uwaoma
*Center for Information Systems & Technology, Claremont Graduate University, 130 E 9th Street, Claremont, CA, USA*
*chinazunwa.uwaoma@cgu.edu*

Keywords: Physical Activity, Smartphone, Respiratory Health, Signal Magnitude Area, Ambient Conditions.

Abstract: While physical activity has been described as a primary prevention against chronic diseases, strenuous physical exertion under adverse ambient conditions has also been reported as a major contributor to exacerbation of chronic respiratory conditions. Maintaining a balance by monitoring the type and the level of physical activities of affected individuals, could help in reducing the cost and burden of managing respiratory ailments. This paper explores the potentiality of motion sensors in Smartphones to estimate physical activity thresholds that could trigger symptoms of exercise-induced respiratory conditions (EiRCs). The focus is on the extraction of measurements from the embedded motion sensors to determine the activity level and the type of activity that is tolerable to individual's respiratory health. The calculations are based on the correlation between Signal Magnitude Area (SMA) and Energy Expenditure (EE). We also consider the effect of changes in the ambient conditions – temperature and humidity, as contributing factors to respiratory distress during physical exercise. Real-time data collected from healthy individuals were used to demonstrate the potentiality of a mobile phone as tool to regulate the level of physical activities of individuals with EiRCs. We describe a practical situation where the experimental outcomes can be applied to promote good respiratory health.

## 1 INTRODUCTION

The importance of physical activity in promoting good health cannot be overemphasized. Rehabilitation therapy for most chronic diseases in recent times, focuses on physical activity training protocols with proven efficacy. Nevertheless, high intensity exercise performed in unfavourable environments is also known to be a common trigger of respiratory conditions such as exercise-induced asthma (EIA), exercise-induced bronchoconstriction or bronchospasm (EIB), exercise-induced rhinitis (EIR), and vocal cord dysfunction (VCD) as reported in these studies (Nielsen et al., 2013; Sinha and David, 2003).

There is a limited capacity to which the respiratory system can adapt to strenuous physical training. The phenomena that affect maximum pulmonary functioning include bronchoconstriction which occurs in the intrathoracic airways and the obstruction of the upper airways (extrathoracic) during intense exercise (Bussotti et al., 2014). And despite the many benefits of regular physical exercise, it is paradoxical to note that those considered to be "highly fit individuals" are more susceptible to respiratory limitation or distress. This technically implies that chronic exercise training results in "faulty adaptation of the lung components" which negatively affects the respiratory system (Bussotti et al., 2014, 48). Where such situation persists, the increased burden and cost of managing these conditions impact the limited health facilities, coupled with shortages of health professionals particularly in remote areas (Braman, 2006; Seto et al., 2009; Surana et al., 2008). Hence, the need to maintain a right balance by applying preventive strategies through consistent monitoring of the level of exercise and the prevailing ambient conditions that could predispose such ailments.

Though several research efforts have concentrated on the use of accelerometer sensors for activity recognition and EE measurements, not so much attention has been given to the type and the level of physical activity appropriate for specific health conditions. This paper focuses on leveraging measurements from motion sensors in smartphones to determine the activity threshold suitable for persons living with EiRCs. We also discuss how variations in environmental condition can affect physical activity level and respiratory health of persons with EiRCs. The key contributions provided by this study are as follows:

(i) We explore and extend the relationship between signal magnitude area and energy expenditure to determine the level of physical activity of an individual in real-time. We consider our approach more convenient and practically simple as it uses only measurements from smartphone motion sensors compared to other methods in previous studies which require using external sensors and gadgets to estimate energy expenditure with the attendant practical limitations.

(ii) We include real-time measurements of variations in the ambient conditions which are noted to have significant impact on the level of engagement of prolonged physical exercise by persons with EiRCs.

## 2 RELATED WORK

Rigorous physical activity has been observed as a major contributor to asthma exacerbation (Bussotti et al., 2014). It has been observed that asthmatics with chronic conditions manifest signs of an asthma attack during exercise (Del Giacco et al., 2015). However, there are many people without asthma who develop symptoms only during such exercises like sporting activities. Milgrom and Taussig (1999) also report that EIA has higher prevalence among children and young adults because of their inclinations to participate in vigorous activities. Though rare, fatal events such as unexpected death during sports reportedly occur more frequently among younger professional athletes than the older ones (Bussotti et al., 2014). A fact sheet from Minnesota Centre for Health Statistics (2004) corroborates these reports where the statistics has it that about 33% of identified school related asthma deaths between 1990 and 2002 occurred during sporting events or Physical Education (PE) Class.

Symptoms of VCD and rhinitis also worsen with highly intensive physical activities performed under adverse environmental conditions (Bussotti et al., 2014). It is important to emphasize here that the severity of EiRCs is related to the type and duration of the physical exertion, as well as the prevailing ambient conditions under which such exercises or activities are performed. Studies in (Bussotti et al., 2014; Del Giacco et al., 2015), observe that persons with EiRCs have airways that are very sensitive to changes in temperature, humidity, and altitude.

Nonetheless, medical experts have advised that respiratory conditions arising from intense physical activities, should not be a justification to deny people an active and healthy lifestyle. Affected individuals can equally participate in sports competitions if such conditions are properly managed and controlled (Milgrom and Taussig, 1999; Bussotti et al., 2014; Del Giacco et al., 2015). The management and treatment of respiratory disorders arising during exercise can benefit from real-time and continuous monitoring; given the social, emotional and economic impact of these health conditions on active and competitive individuals as well as the general populace (Bussotti, Marco, and Marchese 2014; Randolph 1997; Newsham et al. 2002; Keles, 2002).

Studies in recent times employ accelerometer-based measurements for recognition and classification of various physical activities that can provide vital information about individual's health and functional ability (Casamassima et al., 2014; Kwapisz et al., 2011; Chung et al., 2008). Some researchers have also worked on the use of wearable sensors and heart rate monitors for activity recognition and EE estimation during non-steady states and transitions (Altini et al., 2015; Park et al.,2017). In our study, linear motion extracted from the accelerometer measurement is used to determine the patient's level of exercise or physical activity. We also included other motion sensors (gyroscope and digital compass) facilitated by a sensor fusion technique; to generate more intelligent and useful information such as postural changes which may not be provided by using only accelerometer sensor in real-time scenarios. In addition, we used the embedded hygrometer and thermometer in smartphone to record changes in ambient conditions which could contribute to respiratory distress during intense physical activity. By placing the mobile phone securely and strategically on the body trunk, the monitoring system transforms the device motion to body motion. Recordings from smartphone sensors are computed and analyzed to measure deviation from the normal baseline of measured quantities.

## 3 MEASUREMENTS FROM SMARTPHONE SENSORS

### 3.1 Monitoring Physical Activity Level

Monitoring physical activity is recommended as a therapy or rehabilitation approach for persons recovering from cardiac and other related diseases (Kwapisz et al. 2011; Chung et al., 2008). Authors in (Karunanithi et al., 2009) observed that assessment of physical activity of the patient is often by self-reporting, diary, and a 6-minute walk test (6MWT) mostly performed at the hospital exercise clinic or laboratories, using the traditional measure of Metabolic Equivalent Task (MET). Karunanithi et al. (2009) however, proposed an alternative approach to measure physical activity for a home-based care of convalescing patients, where they used MET estimates derived from accelerometer monitors for assessing patients' 6MWT.

The advantages of using an accelerometer sensor for objective measurement and estimation of energy cost by movement intensity include its small size, portability and low power utilization. In (Karunanithi et al., 2009), the authors demonstrate that MET can be derived from accelerometer data which has a high linear correlation ($r^2 = 0.88$) with energy expenditure derived from simultaneous measurement of oxygen consumption ($VO_2$) via conventional gas analyzer. The accelerometer measurement is defined by the signal magnitude area (SMA) in equation (1) while the oxygen consumption is measured by the breath gas analysis. The relationship between the two quantities is shown in equation (2) (Karunanithi et al., 2009).

$$SMA = \int (AccX_{(t)} + AccY_{(t)} + AccZ_{(t)}) \quad (1)$$

$$VO_2 = 1.1 * SMA + 5.7 \quad (2)$$

MET values are estimated from the regression model by averaging the acceleration measurement and oxygen consumption over a given period of time. The intensity of physical activities ranges from sedentary ($\leq 1.5$ METs) to vigorous ($> 6$ METs). 1 MET is equivalent to oxygen uptake while sitting "quietly" (Chuang et al., 2013; Compendium of Physical Activities, 2018).

In our study, the basic metric used in modelling the physical activity level is the movement intensity which measures the instantaneous movement obtained from acceleration signals of the smartphone's built-in accelerometer. However, since the measurement is to be obtained over a given interval rather than momentary, we considered SMA which provides an approximate measurement of energy cost. SMA has been extensively used in previous studies for two purposes – tracking or predicting energy expenditure and discriminating between active and resting states (Chung et al., 2008; Karunanithi et al., 2009). We adopted the SMA metric for the categorization of the physical activity levels because studies have shown that it has a high linear correlation with metabolic rate (MET value), which is widely accepted as a standard measure for movement intensity and energy expenditure (Chuang et al., 2013). Table 1 shows empirically estimated SMA values based on this relationship as documented in Compendium of Physical Activities (2018).

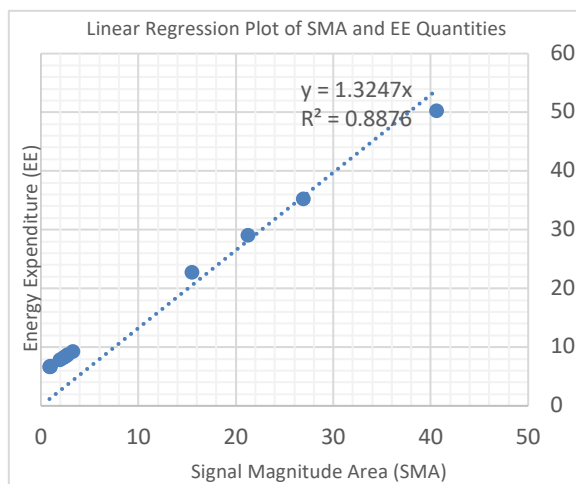

Figure 1: Linear regression and correlation of extrapolated EE values from SMA measurements in the study.

The correlation between EE and SMA values obtained from our study is shown in Figure 1. This is a replication of the study outcome in (Karunanithi et al., 2009).

Monitoring the level of physical activity of persons with EiRCs becomes necessary due to the fundamental role played by vigorous exercise in the inflammation of the lung airways (Del Giacco et al., 2015), which largely contributes to the exacerbation of these conditions. Consistent monitoring of the physical activity level of the affected persons particularly during adverse weather conditions, will help them maintain their active lifestyle and to also participate in top sporting competitions. Interestingly, moderate exercise is being recommended "as a potential therapeutic tool" for persons with EiRCs (Del Giacco et al., 2015).

Table 1: Empirically Estimated SMA Values.

| SMA Range | Designated Activity Level | Examples of Activity Type |
|---|---|---|
| 0.0 - 1.50 | Sedentary | Sitting, Standing, Lying |
| 1.51 – 9.0 | Low | Walking, Yoga, Shopping, etc. |
| 9.1-18.0 | Moderate | Jogging, Hiking, Climbing Stair Case, Callisthenics, etc. |
| Above 18.0 | Vigorous | Bicycling, Running, Treadmill, etc. |

## 3.2 Capturing Ambient Data

Typical circumstances that can accentuate respiratory distress during vigorous exercise are the type and level of physical activity as well as the prevailing environmental conditions. It has been reported that changes in ambient temperature and relative humidity affect the health of persons with EiRCs. Ideal ambient temperature and relative humidity for respiratory health is rated between: 69°F and 79°F – temperature; and 35% and 50% - relative humidity (Venta, n.d; BREATHE- the lung association, n.d).

Breathing in cold and dry air during heavy exercise poses a risk for the development of symptoms of EiRCs (Bussotti et al., 2014; Del Giacco et al., 2015). Embedded ambient sensors in modern mobile phones now provide direct measurement of changes in ambient temperature and humidity. Such information can be used to determine conducive weather for engaging in vigorous and prolonged physical training. In our study, the real-time capturing of ambient condition – temperature and relative humidity on a smartphone is displayed on the chart in Figure 2.

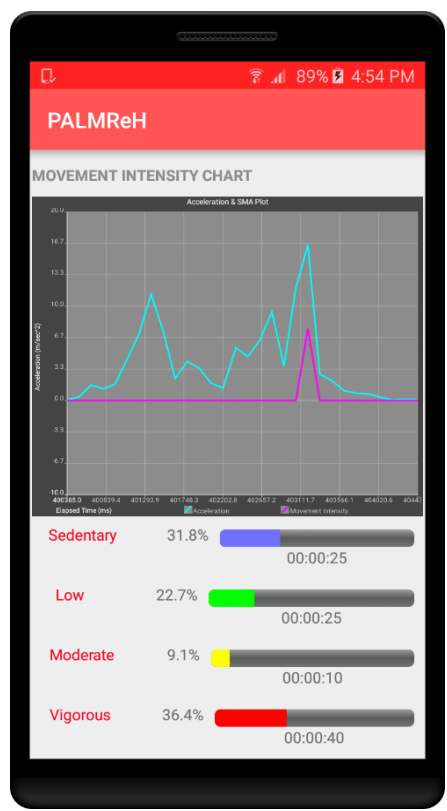

Figure 3: Real-time detection and classification of physical activity levels by the monitoring device.

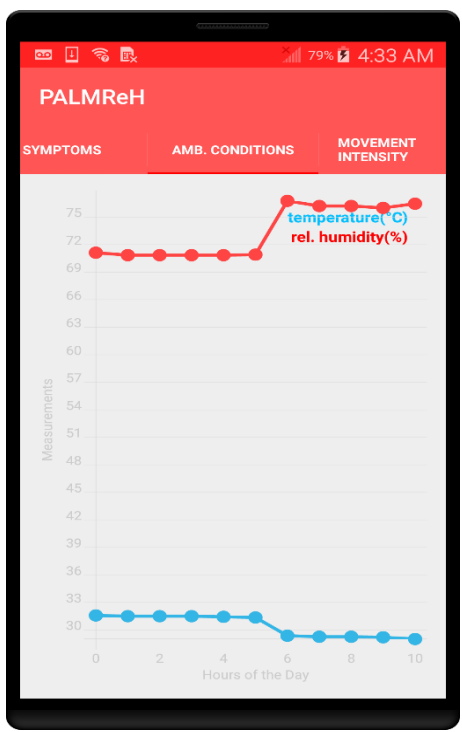

Figure 2: Measurements of ambient temperature and relative humidity captured by the monitoring device.

## 4 EXPERIMENTAL OUTCOMES

In this section, we describe the implementation of the physical activity monitoring on a smartphone. Two quantities of interest in the activity recognition are estimated using the motion sensors. These quantities are *posture change* obtained from the orientation measurement and *activity level* provided by the linear motion parameter. The architecture and the analytical models for these two quantities are described in our previous work (Uwaoma and Mansingh, 2014; Uwaoma and Mansingh, 2018b). The linear movement is represented by the readings of the three accelerometer axes as follows:

- z-axis: captures forward movement;
- y-axis: captures upward/downward movement and
- x-axis: captures horizontal movement.

Table 2: Samples of recorded SMA values extrapolated to EE equivalent with designated Activity Levels.

| SMA (ms$^{-2}$) | EE (VO$_2$) | Activity Level |
|---|---|---|
| 0.986008 | 6.784609 | Sedentary |
| 1.94435 | 7.838785 | Sedentary |
| 0.879925 | 6.667917 | Sedentary |
| 15.5243 | 22.77673 | Moderate |
| 40.58631 | 50.34494 | Vigorous |
| 26.91364 | 35.305 | Vigorous |
| 21.22344 | 29.04579 | Vigorous |
| 2.663409 | 8.62975 | Low |
| 0.935883 | 6.729471 | Sedentary |
| 2.273131 | 8.200444 | Low |
| 3.30391 | 9.3343 | Low |
| 2.463069 | 8.409376 | Low |
| 2.772076 | 8.749284 | Low |
| 1.191858 | 7.011044 | Sedentary |
| 0.69416 | 6.463576 | Sedentary |
| 0.795958 | 6.575553 | Sedentary |
| 2.134268 | 8.047695 | Low |
| 2.250499 | 8.175549 | Low |
| 1.116092 | 6.927701 | Sedentary |
| 2.974935 | 8.972429 | Low |
| 2.019332 | 7.921265 | Low |
| 0.973806 | 6.771186 | Sedentary |
| 1.028279 | 6.831107 | Sedentary |
| 0.338428 | 6.072271 | Sedentary |
| 0.39571 | 6.135281 | Sedentary |
| 0.464635 | 6.211099 | Sedentary |
| 1.649994 | 7.514993 | Low |
| 2.114025 | 8.025427 | Low |

In the experimental test that was performed on the Android platform, we used equation (1) for calculating the SMA values based on the periodical readings on each of the accelerometer's axes. The sampling period was 5000ms which produced 25 samples based on the DELAY_NORMAL accelerometer sampling rate for android devices. To eliminate the noise associated with accelerometer sensor due to its gravity component, a high-pass filter was applied to extract the linear acceleration which was the quantity used to compute the SMA values. Table 2 shows samples of SMA values recorded in real time.

For posture change, the orientation or attitude data is measured in degrees (between 0º and 180º) which is derived from the fused orientation data of the three sensors- gyroscope, accelerometer and magnetometer. Evaluating the posture of the trunk after the user has undergone a burst of energetic or vigorous exercise helps to ascertain if there are remarkable degree of posture variations in terms of the user bending forward or tilting sideways in order to get sufficient air in the event of airway obstruction. The activity level recognition algorithms have been implemented in real life as shown in Figures 3 and 4. The y-axis in Figure 3 displays the acceleration measured in meter per second squared (ms$^{-2}$), while the x-axis displays time in milliseconds. The line plot in magenta represents the SMA for a given period.

Below is a pseudocode implemented in Android which was used to generate the information displayed in Figure 3:

```
Set up a counter for each activity
level detected

Initialize each of the counters to zero

Initialize the total duration in
(ms)for each activity level to zero

Define the sample_period (Android
sampling rate for DELAY_NORMAL is
assumed here)

Set the current activity level to an
empty string

Compute the SMA value for each sampling
period

//Compare the SMA values with the
defined ranges for each activity level

if (SMA value > 0.0 and <= 1.5):
    Increment sedentary_counter
    sedentary_total_duration =
    sedentary_counter * sample_period
    Convert sedentary_total_duration
    to (HH:MM:SS)
    Set current_activity_level to
    "Sedentary"

 else if (SMA value > 1.5 and <= 9.0):
    Increment low_counter
    low_total_duration = low_counter *
    sample_period
    Convert low_total_duration to
    (HH:MM:SS)
    Set current_activity_level to
    "Low"

else if (SMA value > 9.0 and <= 18.00):
    Increment moderate_counter
    moderate_total_duration =
    moderate_counter * sample_period
    Convert moderate_total_duration to
    (HH:MM:SS)
    Set current_activity_level to
    "Moderate"
```

```
else if (SMA value > 18.00):
    Increment vigorous_counter
    vigorous_total_duration =
    vigorous_counter * sample_period
    Convert vigorous_total_duration to
    (HH:MM:SS)
    Set current_activity_level to
    "Vigorous"

Return current_activity_level
```

EiRCs such as EIA and EIB are associated with shortness of breath. For instance, it has been observed that person experiencing asthma distress tends to lean forward in an effort to get sufficient air into the lungs; which invariably makes the person assume an inclined position (Uwaoma and Mansingh, 2018b; WebMD, n.d.).The implementation of the algorithm for postural changes is yet to be tested in real-time; however, we modelled the outcomes graphically as shown in Figure 5. In the model we used the z- axis orientation or tilt angle from the sensor fusion to classify bodily position into four categories:

- Upright position
- Leaning
- Lying
- Inverted

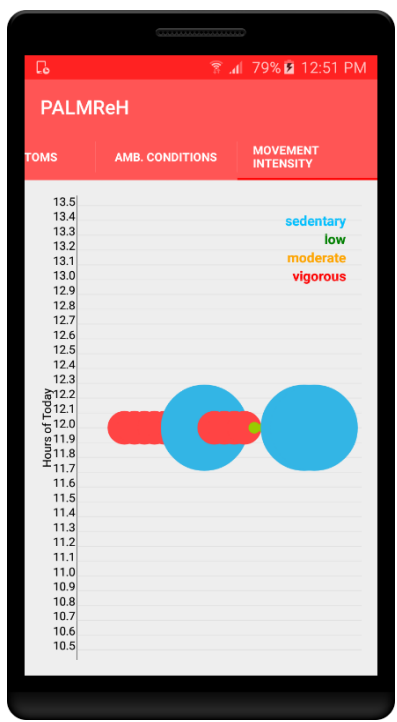

Figure 4: A bubble chart displaying the intensity aggregate of each activity level at given hours of the day.

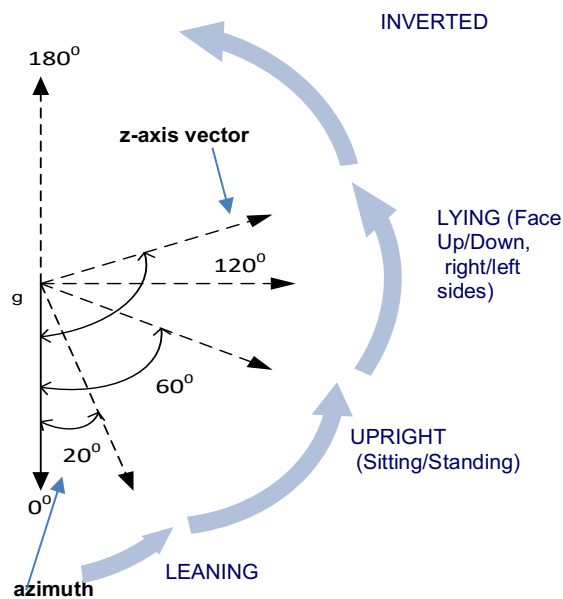

Figure 5: Graphical Modelling of Postural Change.

## 5 SITUATIONAL APPLICATION OF THE PRELIMINARY RESULTS

Here we illustrate a typical practical situation where the preliminary results from the experimental tests can be applied. Asthma persistence and severity are reported to be associated with weight gain in both the paediatric and adult population. (Boulet and Cormiers, 2007, American Lung Association, 2016). Given a scenario where an individual has been diagnosed with obesity and one of the recommended therapies by the physician is regular body exercise; yet, this individual is also diagnosed with a respiratory condition – EIA. Two questions that beg for answers are: Would the patient discountenance the recommended therapy so as to avoid aggravating the respiratory condition? or would the patient engage in prolong body training despite his respiratory condition so as to lose weight as fast possible? The answer to both questions is NO. How to maintain a balance between the two options is where continuous monitoring of the individual's level of physical activity comes into play.

Using the proposed framework, the monitoring device can identify the type and level of physical activity, posture changes, as well variations in the ambient conditions that are not conducive for the individual's respiratory condition, and then signal warnings for the person to take necessary actions to avert exacerbation of his or her condition. Comprehensive details on how the monitoring tool would work to provide such vital information is described in (Uwaoma and Mansingh, 2018a).

# 6 CONCLUSIONS

In this study, we described a framework for determining physical activity threshold for respiratory health, particularly for persons living with EiRCs. We demonstrate how smartphones can be configured to provide a user with vital information with respect to his activity level while engaging in a physical exercise as well as changes in ambient conditions that may contribute to the exacerbation of respiratory distress during physical activity. The major advances here include the ability of the proposed system to concurrently capture the two emphasized measurements – physical activity level and variations in the environmental parameters, benchmarked on gold standards in the study domain. To the best of our knowledge, we are yet to find related work that have considered this approach. However, the focus was on maintaining a balance between engaging in regular physical exercises and managing respiratory ailments that may result from such exercises. This is an on-going study and in our future work, we hope to incorporate measurements like respiratory rate which is a known useful metric for determining a person's respiratory health status.